\documentclass[twocolumn,prl,epsfig,aps]{revtex4}

\usepackage{epsfig}

\begin{document}
\textheight 237mm

\title{\Large Theoretical study of metal borides stability}
\author{Aleksey N. Kolmogorov and Stefano Curtarolo}
\affiliation{
Department of Mechanical Engineering and Materials Science, Duke University, Durham, NC 27708
}

\date{\today}

\begin{abstract}
We have recently identified metal-sandwich (MS) crystal structures and
shown with {\it ab initio} calculations that the MS lithium monoboride
phases are favored over the known stoichiometric ones under
hydrostatic pressure [Phys. Rev. B 73, 180501(R) (2006)]. According to
previous studies synthesized lithium monoboride tends to be
boron-deficient, however the mechanism leading to this phenomenon is
not fully understood. We propose a simple model that explains the
experimentally observed off-stoichiometry and show that compared to
such boron-deficient phases the MS-LiB compounds still have lower
formation enthalpy under high pressures. We also investigate stability
of MS phases for a large class of metal borides. Our {\it ab initio}
results suggest that MS noble metal borides are less unstable than the
corresponding AlB$_2$-type phases but not stable enough to form under
equilibrium conditions.

\end{abstract}

\maketitle

\section{1. Introduction}
\label{section.introduction}

The interest in the AlB$_2$ family of metal diborides re-emerged after
the discovery of superconductivity in MgB$_2$ with a surprisingly high
transition temperature of 39 K\cite{origin}. Boron $p$-states have
been shown to be key for both stability and superconductivity in these
compounds\cite{Kortus,Shein,Oguchi}. MgB$_2$ is a unique metal
diboride because it has a significant density of boron
$p\sigma$-states at the Fermi level which give rise to the high T$_c$
superconductivity, and yet enough of them are filled for the compound
to be structurally stable\cite{Kortus,Shein,Oguchi}. The effectively
hole-doped noble- and alkali-metal diborides would have higher
$p\sigma$ density of states (DOS) at E$_F$, but they have been
demonstrated to be unstable under normal conditions\cite{Oguchi}. The
effort to achieve higher T$_c$ has thus primarily focused on doping
magnesium diboride with various metals; however, doping this material
has proven to be difficult\cite{dope_review} and no improvement on
T$_c$ has yet been reported. According to a recent theoretical study
of nonlocal screening effects in metals, MgB$_2$ may already be
optimally doped\cite{peihong}. Lithium borocarbide with a doubled
AlB$_2$ unit cell has been suggested as a possible high-T$_c$
superconductor under hole-doping\cite{LiBC}, but disorder in the
heavily doped Li$_x$BC appears to forbid superconductivity above
2K\cite{LixBC}.

In this work we investigate whether there could be stable high-T$_c$
superconducting metal borides in configurations beyond the standard
AlB$_2$ prototype. We have recently proposed metal-sandwich (MS)
structures MS1 and MS2, which also have $sp^2$ layers of boron but
separated by two metal layers\cite{MGB}. Despite their rather simple
unit cells these structures have apparently never been considered
before. As we demonstrate below, identification of the MS structures is
not straightforward because they represent a local minimum not usually
explored with current compound prediction strategies\cite{MGB}. We
reveal trends in the cohesion of MS phases by calculating formation
energies for a large class of metal borides and show that some
monovalent-metal borides benefit from having additional layers of
metal. The MS noble-metal borides still have positive formation
energy, but they are less unstable than the AlB$_2$-type phases. This
result helps resolve the question of what phases would form first in
the noble-metal boride systems under non-equilibrium
conditions\cite{AgxB2,hype,PF,Ag_laser}.

Our main finding concerns the Li-B system, in which the MS lithium
monoboride is stable enough to compete against the known
stoichiometric phases. According to our previous {\it ab initio}
calculations the MS lithium monoboride is comparable in energy to
these phases under normal conditions, but it becomes the ground state
at 50\% concentration under moderate hydrostatic
pressures\cite{MGB}. Here we extend the analysis to non-stoichiometric
Li-B phases which could potentially intervene in the synthesis of the
MS phases. In particular, synthesized lithium monoboride with linear
chains of boron is known to be boron-deficient for reasons not fully
understood so far. We simulate the incommensurate LiB$_y$ phases
(notation explained in Ref. \cite{xy}) by constructing a series of
small periodic Li$_{2n}$B$_m$ structures and show that the minimum
formation energy is achieved for $y\approx0.9$, in very good agreement
with the experimentally observed values. Using this simple model of
the off-stoichiometry phases with linear chains of boron we
demonstrate that relative to them MS-LiB still has lower formation
enthalpy under high pressures. Simulations of other alkali-metal
borides, MB$_y$ (M = Na, K, Rb, Cs), suggest that that these nearly
stoichiometric phases might form under moderate pressures.

The paper is divided in the sections describing: 2) simulation details;
3) construction of the MS prototypes; 4) stability of MS phases for a
large class of metal borides; 5) detailed investigation of the Li-B
system; 6) simulations of other monovalent and higher-valent metal
borides; 7) summary of the electronic and structural properties of the
MS phases.

\section{2. Computation details}
\label{section.methods}

Present {\it ab initio} calculations are performed with Vienna
Ab-Initio Simulation Package {\small VASP}
\cite{kresse1993,kresse1996b} with Projector Augmented Waves
(PAW)~\cite{bloechl994} and exchange-correlation functionals as
parametrized by Perdew, Burke, and Ernzerhof (PBE)\cite{PBE} for the
Generalized Gradient Approximation (GGA). Because of a significant
charge transfer between metal and boron in most structures considered
we use PAW pseudopotentials in which semi-core states are treated as
valence. This is especially important for the Li-B system as discussed
in Refs. \cite{MGB,US_PAW}. Simulations are carried out at zero
temperature and without zero-point motion; spin polarization is used
only for magnetic alloys. We use an energy cutoff of 398 eV and at
least 8000/(number of atoms in unit cell) ${\bf k}$-points
distributed on a Monkhorst-Pack mesh \cite{MONKHORST_PACK}.  We also
employ an augmented plane-wave+local orbitals (APW+lo) code {\small
WIEN2K} to plot characters of electronic bands\cite{WIEN2K}. All
structures are fully relaxed. Our careful tests show that the relative
energies are numerically converged to within 1$\sim$2 meV/atom.

Construction of binary phase diagrams A$_x$B$_{1-x}$ is based on the
calculated formation enthalpy $H_{f}$, which is determined with
respect to the most stable structures of pure elements. For boron
there are two competing phases $\alpha$-B and $\beta$-B\cite{bB}; we
use $\alpha$-B (Ref. \cite{Oguchi}), theoretically shown to be the
more stable phase at low temperatures and high pressures \cite{bB}. A
structure at a given composition $x$ is considered stable (at zero
temperature and without zero-point motion) if it has the lowest
formation enthalpy for any structure at this composition and if on the
binary phase diagram $H_{f}(x)$ it lies below a {\it tie-line}
connecting the two stable structures closest in composition to $x$ on
each side.

\section{3. Identification of MS prototypes}
\label{section.identification}

Data-mining of quantum calculations (DMQC), introduced in our previous
work\cite{SC1}, is a theoretical method to predict the structure of
materials through efficient re-use of {\it ab initio} results. The
DMQC iteratively determines correlations in the calculated energies on
a chosen library of binary alloys and structure types. The last work
has demonstrated that for a set of 114 crystal structures and 55
binary metallic alloys the method gives an almost perfect prediction
of the ground states (within the library) in a fraction of all
possible computations\cite{SC1,Morgan}. The speed-up (commonly by a
factor from 3 to 4) is achieved by the method's rational strategy for
suggesting the next phase to be evaluated. An essential feature of
these calculations is the full relaxation of the considered
structures, which ensures an accurate determination of the
correlations in the chosen library\cite{DMQC}.

We have recently begun expanding the $114\times55$ library of {\it ab
initio} energies of binary alloys\cite{SC1} to include metal borides.
Boron tends to form covalent bonds in intermetallic compounds; to have
this correlation in future predictions with the DMQC we needed first
to add a few compound-forming metal-boride systems into the
library. Introduction of a new system involves calculations of
energies for all the prototype entries in the library. Surprisingly,
in the very first system considered, Mg-B, one of the fcc structures
with 4-atom unit cell at 50\% concentration, A$_2$B$_2$ fcc-(111) (or
V2 \cite{Zunger1}), relaxed almost all the way down to the
AlB$_2$-MgB$_2$$\leftrightarrow$hcp-Mg tie-line. Significant
relaxations are not uncommon in our simulations; they usually
correspond to the transformation from a starting configuration to a
known stable prototype and are automatically detected by the change in
the symmetry. The magnesium monoboride phase, however, retained its
original space group R$\bar{3}$m (\#166).
\begin{figure}[t]
  \begin{center}
    \centerline{\epsfig{file=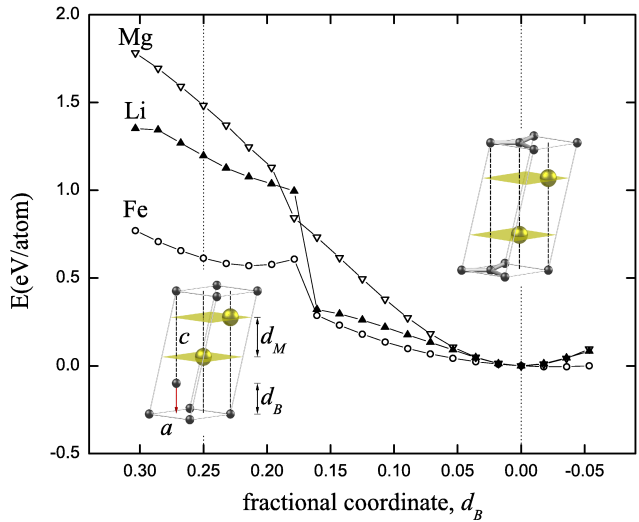,width=95mm,clip=}}
    \vspace{-1mm} \caption{\small (color online). Energy barrier as a
    function of boron fractional coordinate $d_B$ for a transition
    path from V2 ($d_B\approx 0.25$ \cite{Zunger1}) to MS1
    ($d_B\approx 0$, see text) for different metal borides. For each
    value of $d_B$ the energy is minimized by adjustment of the
    remaining free parameters $a$, $c$, and the metal fractional
    coordinate $d_M$.}
\label{orig}
\end{center}
\end{figure}

Having examined the relaxation process we found that there is a
continuous symmetry-conserving path from V2 to a new structure
MS1\cite{MGB,MS1}. V2 has 4 atoms per unit cell with 4 free parameters
$a$, $c$, $d_B$, and $d_M$ (the last two are fractional distances
between boron and metal layers), so that atoms are constrained only to
the vertical lines (see Fig. 1). The perfect fcc lattice corresponds
to $c/a$=4$\sqrt{2/3}$ and $d_B=d_M$=0.25, but this special case does
not grant additional symmetry operations and local relaxation have
been seen in some metallic systems\cite{Zunger1}. In metal borides a
more dramatic transformation leads to a much more stable
configuration: boron atoms rearrange themselves to form covalent bonds
in a hexagonal layer ($d_B\rightarrow 0$) rather then share electrons
in close-packed triangular layers, while metal atoms remain in a
close-packed bilayer. We have checked other alkali, alkaline and
transition metal borides not present in the DMQC library and confirmed
that they all benefit from this transformation; however, some
electron-rich systems might not escape from the local fcc-type
minimum, as shown in Fig. 1 for FeB. This could be a reason why the
MS1 prototype has apparently been overlooked so far. We would like to
point out that identification of new prototypes is not an intended
function of the DMQC. This interesting accidental result should be
credited to the exhaustive consideration of all candidates (regardless
of how unlikely they seem to be a stable phase - an fcc supercell is
hardly a suitable configuration for a magnesium boride phase) and the
careful structural relaxation in the calculation of their ground state
energies.

We proceed by constructing a library of related MS prototypes.
Structures, where the metal atoms closest to boron sit directly above
the center of boron hexagons, are uniquely specified by the positions
of the metal layers (such as $\alpha$, $\beta$, $\gamma$, in Fig. 2).
The MS1 structure can thus also be labeled as $|\alpha\beta|$: the
Greek letters show the positions of the two metal layers and vertical
bars correspond to boron layers. A hexagonal supercell for this
phase\cite{MS1} is obtained when the last metal layer matches the
first: $|\alpha\beta|\beta\gamma|\gamma\alpha|$. The fourth metal
layer can alternatively be shifted back to site $\alpha$ (see
Fig. 2(d)), resulting in another structure at the same stoichiometry
MS2 ($|\alpha\beta|\beta\alpha|$)\cite{MGB,MS2}. Examples of more
metal-rich structures are $|\alpha\beta\alpha|$ (MS3)\cite{MS3} and
$|\alpha\beta\gamma|$.  Various stoichiometries can also be achieved
by combination of smaller cells, i.e. $|\alpha\beta|\beta|$ (MS4) and
$|\alpha|\alpha\beta|\beta\alpha|$. In this notation the
AlB$_2$-prototype is labeled simply as $|\alpha|$, which we will use
henceforth to avoid confusion with the aluminum diboride compound.
Positioning metal atoms above boron hexagon centers is not the only
possibility. Stacking faults have been experimentally observed in
MgB$_2$\cite{stack_exp}, though such defects have been shown to be
energetically costly for this compound\cite{stack_theory}. We have
constructed a periodic structure ($\delta$-MB$_2$) with 3 atoms per
unit cell where the metal atoms in $|\alpha|$ are shifted along ({\bf
a}+{\bf b}) to be above the middle of a boron-boron
bond\cite{deltaAu}.
\begin{figure}[t]
  \begin{center}
    \centerline{\epsfig{file=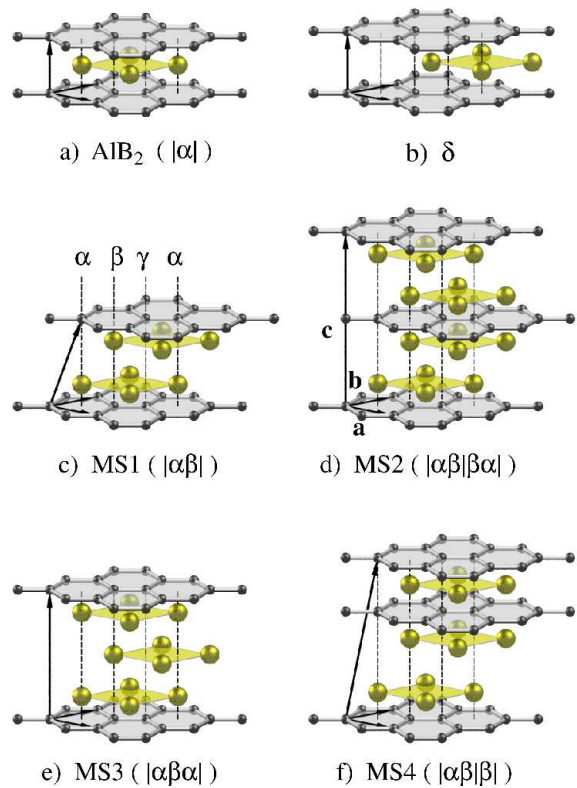,width=75mm,clip=}}
    \vspace{-1mm}
    \caption{\small (color online). Known AlB$_2$ and proposed MS
      structures for metal borides. The hexagonal layers of boron
      (grey) are separated by triangular layers of metal (yellow, only
      four atoms per layer are shown). Except for the $\delta$
      structure shown in b)\cite{deltaAu}, all structures have metal
      layers positioned above the middle of the nearest boron
      hexagons. The alternative notations given in brackets are
      explained in the text. The concentration of metal in these
      structures is 33\% (AlB$_2$), 33\% ($\delta$), 50\% (MS1), 50\%
      (MS2), 60\% (MS3), and 43\% (MS4).}
\label{mgb}
\end{center}
\end{figure}
%
\section{4. Stability of the MS phases}
\label{section.stability}

Formation of a particular compound in systems with a few competing
phases is determined by a number of factors, i.e. the ground state
energy, synthesis conditions, thermodynamic and kinetic
effects. Comparison of high-throughput {\it ab initio}
results\cite{Morgan,CALPHAD,MGA} with experimental databases has shown
that the calculated total energies alone allow to identify the correct
phases observed in the experiment in about 96.7\% of investigated
cases (Eq. 3 in Ref. \cite{CALPHAD}). In this section we use the total
energy criterium to narrow down the set of systems in which the MS
phases might occur.
\begin{figure}[t]
  \begin{center}
    \centerline{\epsfig{file=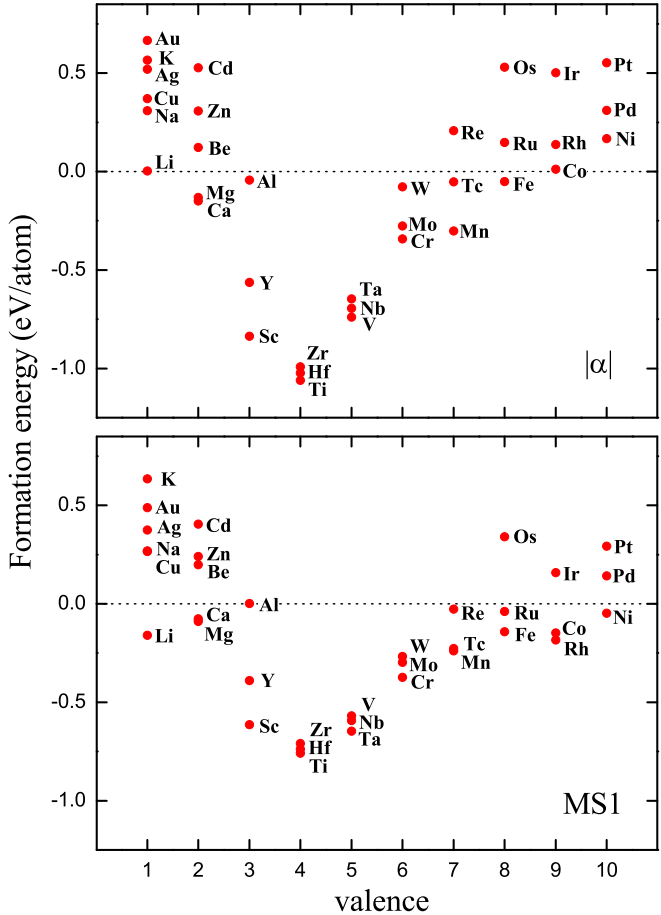,width=97mm,clip=}}
    \caption{\small (color online). Calculated formation energy for
    $|\alpha|$ (top) and MS1 (bottom) metal borides. The results for
    $|\alpha|$-MB$_2$ are consistent with the previous
    study\cite{Oguchi} (mind the minus sign difference between
    formation energy and heat of formation).}
    \label{mb} \end{center}
\end{figure}

We first calculate the formation energy for a large library of alkali,
alkaline, and transition metals in the $|\alpha|$ and MS1
configurations (Fig. \ref{mb}). Our results for $|\alpha|$-MB$_2$ are
consistent with the previous calculation by Oguchi\cite{Oguchi} (note
that we put noble metals in the first valence group). The MS1 phases
exhibit a similar trend in cohesion: they are most stable for
tetra-valent metal borides. It is convenient to analyze the stability
of the MS phases by comparing them to the corresponding
$|\alpha|$-phases because the MS structures are effectively a
combination of the $|\alpha|$-structure and additional layers of
metal. Three immediate effects can be expected from the insertion of
an extra metal layer: 1) different strain conditions between the boron
network and the triangular layers of metal; 2) different doping level
of the boron layer; 3) significant reduction of interlayer overlaps
between $p$-orbitals of boron due to the increase in the interlayer
distance. To decouple these effects we calculate {\it relative
stability} of the MS1 phase with respect to phase separation into
$|\alpha|$ and pure element for the large library of metal
borides. The relative stability for compound M$_x$B$_{1-x}$ ($\equiv X$)
is defined as
\begin{eqnarray}
  \Delta E_{X} \equiv E_{X}-\frac{3}{2}\left[(1-x)E_{|\alpha|}+\left(x-\frac{1}{3}\right)E_{pure}\right]
\end{eqnarray}
(all energies are per atom), and reflects whether a metal layer
prefers to be in a layered boride environment or stay in pure bulk
structure. To illustrate the amount of strain in the system we plot
this energy difference versus equilibrium intralayer distance in pure
fcc or hcp bulk metal structure, whichever is more stable at zero
temperature\cite{bcc}. Figure 4 shows that monoborides of metals in
the same valence group and similar dimensions (for example Zn, Cd, and
Mg) have close relative stability. Metal layers mismatched with the
boron layer (for example Be, Ca, Na, and K) cause a significant energy
penalty when inserted in the respective diborides.
\begin{figure}[t]
  \begin{center}
    \centerline{\epsfig{file=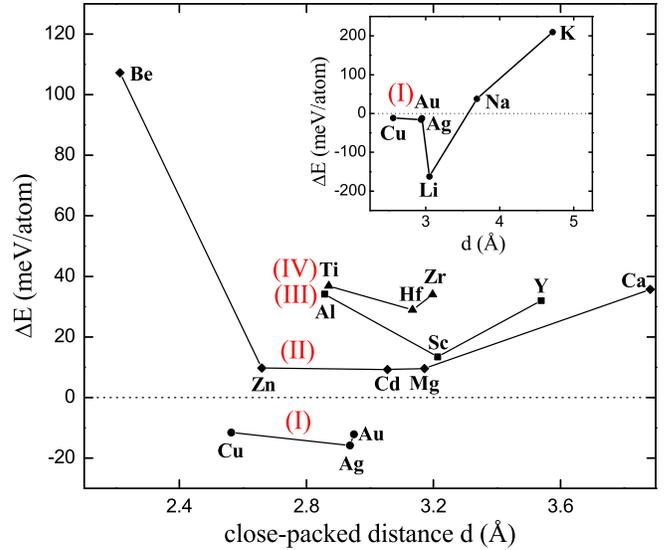,width=97mm,clip=}}
    \caption{ \small (color online). Relative stability of the MS1
      metal borides with respect to phase separation into $|\alpha|$
      and pure metal phases (Eq. 1) as a function of the close-packed
      distance defined in the text. The metal borides are grouped and
      labeled according to the valence of metal (from 1 to 4). Inset:
      the same for noble- and alkali-metal borides.}
      \label{all} \end{center}
\end{figure}

Another general trend captured in Fig. 4 is a consistent decrease of
the MS1 phase relative stability with the increase of valence
electrons (up to four). Relative stability depends on the binding
mechanisms in all the three phases in Eq. 1 and the analysis of its
variation with the metal valence is not straightforward. For
monovalent metal borides the observed gain in binding for the MS1
phase is consistent with the fact that diborides of low-valent metals
have available $p\sigma$ bonding states and stabilize as the metal
valence increases\cite{Kortus,Oguchi}. However, charge redistribution
in these phases may follow different scenarios: in $|\alpha|$ metal
atoms are exposed to boron and become almost entirely
ionized\cite{Kortus}, while in MS1 boron extracts charge through the
surface of the metal bilayer and likely leaves more charge in the
metal system (Section 7). With increasing valence in the transition
metal series of diborides bonding $p\sigma$ and $d$-$p\pi$ bands
become occupied, so that the binding reaches its maximum for Group IV
metals and eventually goes down\cite{Oguchi}. Consequently, we observe
a noticeable increase in the relative stability of the MS1 structure
for higher-valent metals (in fact, all metal diborides with at least
five valence electrons benefit from insertion of an extra metal layer
with the largest gain of -340 meV/atom obtained for RhB). However,
these electron-rich systems allow other phases with significantly
lower energies (prototypes NiAs, NaCl, FeB-b, etc.\cite{PF}). Hence,
we focus on low-electron systems that have been shown to stabilize
through incorporation of extra metal layers and could compete with
existing phases.

\section{5. Li-B system}
\label{section.Li-B}

{\it Overview.} A few compounds at different stoichiometries have been
reported for the Li-B system\cite{PF,aLi,bLi,Alkali,B-Li,Li3B14,LiB3}.
On the boron-rich side the experimentally reproducible compounds
Li$_3$B$_{14}$ and LiB$_3$ have large unit cells with fractional
occupancies\cite{Li3B14,LiB3,PF} and cannot be presently simulated
with {\it ab initio} methods with desired degree of accuracy. The
composition of the most lithium-rich LiB$_y$ compounds (near 50\%
concentration) apparently depends on synthesis conditions and
post-synthesis treatment, as the reported values for $y$ range from
0.8 to 1 (notation explained in Ref. \cite{xy}). In the early
experiments the formed compounds were ascribed compositions
Li$_5$B$_4$\cite{Wan78,Wan79} or Li$_7$B$_6$\cite{Dal79}; Wang {\it at
al.} used a rhombohedral model to explain the observed x-ray
patterns\cite{Wan78,Li5B4}. However, a more consistent interpretation
of the available x-ray data on nearly stoichiometric lithium
monoboride has been recently given by Liu {\it et al.}\cite{aLi}. The
authors demonstrated that the main x-ray peaks can be indexed with a
four-atom hexagonal unit cell $\alpha$-LiB (Fig. 5(a)), which consists
of linear chains of boron embedded in hexagonal lithium
shells\cite{aLi}.

While the simple $\alpha$-LiB sheds light on what the structure of the
lithium monoboride is, an important question remains open as to why
the LiB$_y$ compounds are boron-deficient. W\"{o}rle and Nesper have
offered an insightful model of LiB$_y$, in which the boron chains are
uncorrelated and incommensurate with the lithium
sublattice\cite{Wor00}. By using a large unit cell containing 32,000
atoms the authors reproduced a kink at $2\theta\approx 60^{\circ}$ in
the x-ray pattern and attributed it to the average boron-boron
distance of 1.59 \AA. They also suggested that the boron chains could
be dimerized or have vacancies\cite{Wor00}. According to a recent
theoretical study, boron chains in lithium monoboride are not expected
to dimerize but might indeed be able to slide freely along the lithium
sublattice\cite{bLi}.
\begin{figure}[t]
  \begin{center} \vspace{-5mm}
    \centerline{\epsfig{file=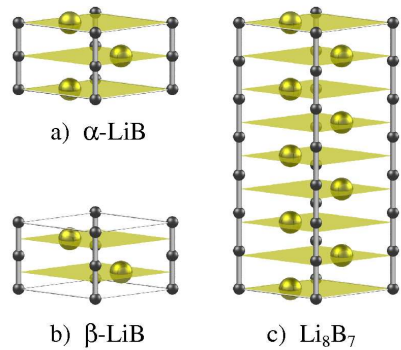,width=60mm,clip=}}
    \caption{ \small (color online). Structures with linearly chained
    boron immersed in a lithium sublattice: the known $\alpha$-LiB and
    $\beta$-LiB have commensurate sublattices; the proposed
    Li$_{2n}$B$_m$ series (only Li$_8$B$_7$ is shown) allows to
    simulate the incommensurate off-stoichiometry LiB$_y$ phases.}
  \label{press} \end{center}
\end{figure}

{\it Model of the LiB$_y$ compounds.} Simulation of the disordered
LiB$_y$ compounds is essential for finding a possible stability region
of the MS-LiB phases. While the large unit cells with thousands of
atoms are needed to reproduce the x-ray data, such sizes prohibit the
use of {\it ab initio} methods for ground state energy
calculations. Therefore, we simulate the incommensurate LiB$_y$
compounds by constructing a series of relatively small commensurate
Li$_{2n}$B$_m$ phases. The number of lithium atoms in a unit cell must
be even since they occupy alternating sites along the $c$-axis. To
determine the optimal relative position of the two sublattices we fix
only the $z$-components of one lithium and one boron atoms and allow
all the other degrees of freedom to relax. We find that the relative
placement matters only for the smallest Li$_2$B$_2$ unit cells: as we
have shown in Ref. ~\cite{MGB} the energy difference between
$\alpha$-LiB and $\beta$-LiB is 10 meV/atom. For all other periodic
structures the barriers to sliding for the two sublattices are below 1
meV per unit cell. The situation is similar to the relative motion in
multiwalled carbon nanotubes, where the rigid layers interact weakly
with one another: in long-period commensurate nanotubes the barriers
to intertube sliding are extremely small, and in incommensurate ones
the intertube sliding mode is gapless\cite{AL}.  Local relaxations in
Li$_{2n}$B$_m$ ($m>2$, $m\not=2n$) are insignificant due to the
rigidity of the boron-boron chains and a smooth charge density
distribution along the chains\cite{bLi}.
\begin{figure}[t]
  \begin{center} \vspace{-5mm}
    \centerline{\epsfig{file=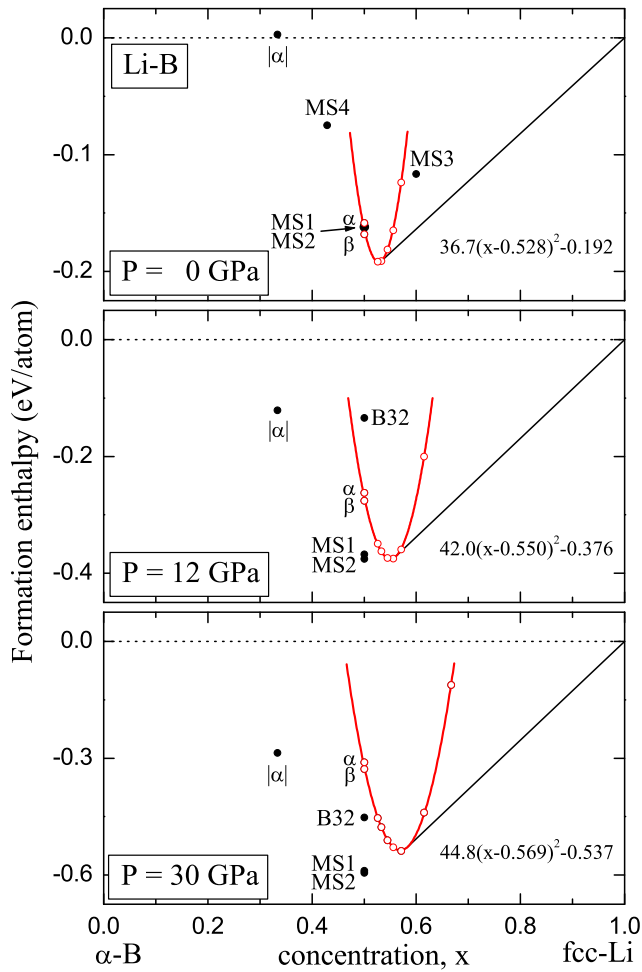,width=95mm,clip=}}
    \vspace{-8mm} \caption{ \small (color online). Calculated phase
    diagrams for the Li-B system. Hollow red points are M$_{2n}$B$_m$
    phases, from left to right: $\alpha$-LiB, $\beta$-LiB,
    Li$_{10}$B$_9$, Li$_8$B$_7$, Li$_6$B$_5$, Li$_{10}$B$_8$,
    Li$_4$B$_3$ for all pressures, Li$_8$B$_5$ for $P$ = 12 GPa, and
    Li$_8$B$_5$, Li$_2$B for $P$ = 30 GPa. The parameters of the
    parabolic fit are given on each panel; the solid lines are tangent
    to the parabolas at $x=0.534$, $x=0.560$, and $x=0.584$ for 0, 12,
    and 30 GPa pressure, respectively. $|\alpha|$-LiB is the AlB$_2$
    prototype and B32 is the pseudodiamond structure\cite{aLi}. Note
    the different enthalpy scale for different pressures.}
    \label{PHD} \end{center}
\end{figure}
\begin{figure}[t]
  \begin{center} \vspace{-5mm}
    \centerline{\epsfig{file=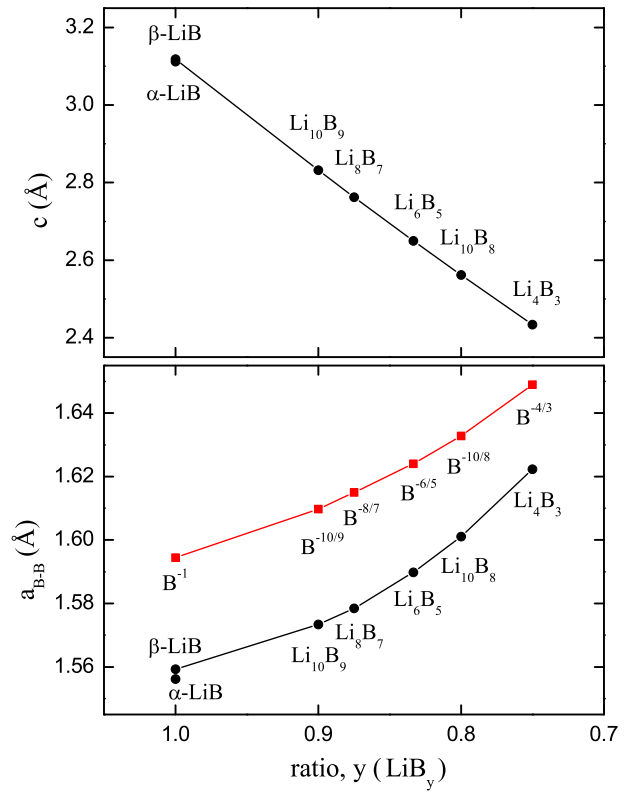,width=90mm,clip=}}
    \vspace{-8mm} \caption{ \small (color online). Calculated
    optimized lattice parameters in phases with linear chains of
    boron. Top panel: $c$-axis distance (the doubled Li-Li interlayer
    spacing, 2$c_{Li-Li}$) as a function of the B to Li ratio in
    LiB$_y$. Bottom panel: the black circles are the B-B bond lengths
    in the LiB$_y$ phases ($a_{B-B}=c/y/2$); the red squares are B-B
    bond lengths in the B$^{-1/y}$ phases explained in the text.}
    \label{axis} \end{center}
\end{figure}

{\it Stability and structure of the LiB$_y$ compounds.} Figure 6 shows
an immediate benefit for the lithium monoboride to change composition:
as the level of lithium concentration increases by a few percent the
phase undergoes stabilization by over 20 meV/atom at zero pressure.
The points on the enthalpy versus concentration plot for the
Li$_{2n}$B$_m$ series nicely fit to a parabola (with minimum at
$y\approx0.894$ at zero pressure). This leads to an interesting
situation, in which the LiB$_y$ phases are actually stable over a {\it
range} of concentrations. The lower boron concentration limit
corresponds to $y_{min}\approx0.874$ (the tangent to the parabola
going through $x=1$, shown in Fig. 6), while the higher one depends on
the location of LiB$_3$ on the phase diagram and should be around
$y_{max}\approx0.9$. The allowed concentrations are in excellent
agreement with the W\"{o}rle and Nesper's value of $y=0.9$ inferred from
the analysis of the x-ray data\cite{Wor00}. Considering that the
lithium and boron sublattices are nearly independent, it seems
possible to manipulate the stoichiometry with active solutions by
removing the alkali metal through the surface of the sample. By using
tetrahydrofuran-naphtalene solution Liu {\it et al.} may have
extracted not only the free lithium, but also the lithium from the
LiB$_y$ compound pushing the concentration of boron towards the higher
limit of the stability region ($y_{min}$,$y_{max}$), and maybe beyond
it.

To help determine the composition of the LiB$_y$ compounds from
experimental data we plot the fully relaxed lattice parameters in
Fig. 7. We observe that the $c$-axis undergoes an almost linear
expansion with the increase of the boron to lithium ratio: $c= 0.365 +
2.746 y$. Because of the 1$\sim$2\% systematic errors in the bond length
calculations within the GGA these results cannot be used to pinpoint
the absolute value of $y$. However, the {\it variation} of the lattice
parameters as a function of $y$ is expected to be much more accurate
and allows one to estimate the range of concentrations for the
experimentally observed compounds. For example, the measured
$c=2.875(2)$ \AA\ and $c=2.792(1)$ \AA\ values corresponded to nominal
compositions $y=1.0$ and $y=0.82$, respectively\cite{Wor00}; the slope
from Fig. 7 indicates that the difference in $y$ in the synthesized
compounds was, in fact, about 6 times smaller ($\Delta y=0.03$). The
discrepancy in the measured (2.796 \AA, Ref. \cite{aLi}) and
calculated (3.102 \AA, Ref. \cite{bLi}) $c$-axis values pointed out by
Rosner and Pickett\cite{bLi} can be explained as that the synthesized
compound was not a stoichiometric lithium monoboride but rather
LiB$_{y\approx 0.89}$.
\begin{figure}[t]
  \begin{center} \vspace{-5mm}
    \centerline{\epsfig{file=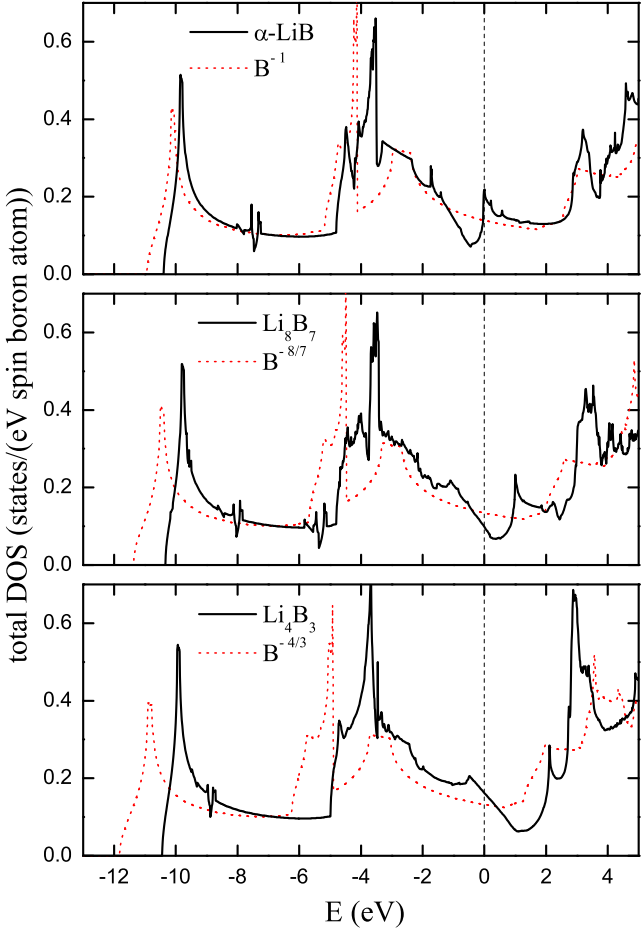,width=95mm,clip=}}
    \vspace{-8mm} \caption{ \small (color online). Calculated total
    density of states (DOS) for the Li$_{2n}$B$_m$ compounds (solid
    black lines) and hypothetical B$^{-1/y}$ phases explained in the
    text (dotted red lines). Fermi level is at 0 eV.}
    \label{tnew} \end{center}
\end{figure}
\begin{figure}[t]
  \begin{center} \vspace{-5mm}
    \centerline{\epsfig{file=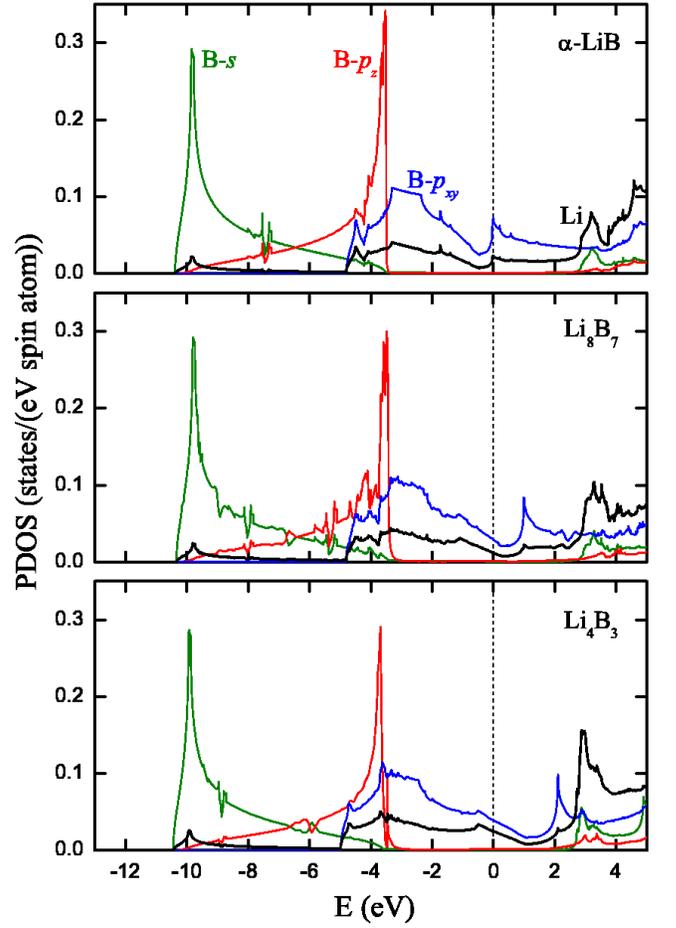,width=95mm,clip=}}
    \vspace{-8mm} \caption{ \small (color online). Calculated partial
    density of states (PDOS) for the Li$_{2n}$B$_m$ compounds. The
    curves correspond to B-$s$ (green), B-$p_z$ (red), B-$p_{xy}$
    (blue), and Li (black) characters. Fermi level is at 0 eV.}
    \label{pdos} \end{center}
\end{figure}

It should be noted that the determination of the concentration
($y=a_{B-B}/c_{Li-Li}$, $c_{Li-Li}=c/2$) presents difficulty only
because the length of the boron-boron bond is hard to extract from the
experiment\cite{Wor00}. However, it is the covalent boron-boron bond,
being very rigid, that determines the $c$-axis dimension. This can be
best illustrated by simulating linear chains of boron in hypothetical
structures B$^{-1/y}$ where lithium is replaced with the equivalent
number of extra electrons $1/y$ (a uniform positive background is used
here to impose charge neutrality). The B$^{-1/y}$ structures with one
boron atom per unit cell keep the basal lattice vectors of the
corresponding LiB$_y$ compounds while the $c$-axis is
optimized. Fig. 7 shows that the charge transfer from lithium to boron
alone can explain the variation of the bond length for
$y=0.8\sim1.0$. The origin of the nearly constant offset between the
two curves becomes evident when the lithium sublattice is, in turn,
simulated without boron. A hypothetical Li$^{+1}$ structure with the
same in-plane dimensions as in $\alpha$-LiB has a much shorter
equilibrium $c$-axis of 2.22 \AA. The $E_{Li^{+1}}(c)$ dependence in
the 2.5$\sim$3.1 \AA\ range of $c$ is almost linear with a slope of
0.76 eV/\AA. As a result, the lithium sublattice in the LiB$_y$ phases
exerts a small stress on the linear chains of boron for all $y$ from 1
to 0.8. The stress induces the shortening of the boron-boron bond by
0.029 \AA\ ($y=1$) and 0.033 \AA\ ($y=0.8$)\cite{stress} and thus
turns out to be the main reason for the 0.035 \AA\ ($y=1$) and 0.032
\AA\ ($y=0.8$) bond length mismatches in the simulated phases with and
without lithium (Fig. 7). The result illustrates why LiB$_y$ can be
represented well as a superposition of the two electronically
independent doped sublattices. However, our next test shows that
lithium does play an important role in defining the optimal
composition of LiB$_y$ by affecting the electronic states of boron
near the Fermi level.

To further investigate the mechanism leading to the existence of the
off-stoichiometry lithium borides we plot the total and partial DOS
for several Li$_{2n}$B$_m$ phases in Figs. 8 and 9 and the band
structure in a representative $\alpha$-LiB phase in Fig. 10. The
states near the Fermi level are hybridized $p\pi$-B and Li
states\cite{bLi}. The average presence of the Li character in the DOS
in the -10 to 3 eV energy range is small in $\alpha$-LiB
($N^{Li}/N^B_{p\pi}\approx 0.3$), but becomes more significant in
Li$_4$B$_3$ ($N^{Li}/N^B_{p\pi}\approx 0.5$). The $p\pi$ boron states
extend into the lithium-filled interstitials the furthest, so they are
affected by the electrostatic potential from the lithium ions the most
(in fact, the van Hove singularity in $\alpha$-LiB at $E=0$ is not
present in the corresponding B$^{-1}$ structure described above, see
dotted curves in Figs. 8 and 10). Note that the bonding and
antibonding boron states are not well separated in the LiB$_y$ phases:
in the case of $\alpha$-LiB both types are present in the 0-1.3 eV
energy range (Fig. 10). Since the Fermi level in $\alpha$-LiB (Fig. 8)
already catches the edge of the antibonding $p\pi$-B states, one would
naturally expect for the compound to benefit from {\it losing}
lithium. However, the system does not follow the rigid band scenario
as the concentration of lithium increases: the van Hove singularity is
pushed away from the Fermi level (in $\alpha$-LiB) to the right by
over 2 eV (in Li$_4$B$_3$) (see Fig. 8). The optimal position of the
Fermi level near the bottom of the $p\pi$ pseudogap is achieved in
Li$_{10}$B$_9$ and Li$_8$B$_7$, which is a part of the reason why the
LiB$_y$ phases have the minimum formation energy at $y\approx
0.9$. The correlation between the position of the Fermi level in the
pseudogap and the maximum stability has been observed in various
systems\cite{Oguchi,EF}.

{\it Comparison of the MS phases with the known metal borides phases
under pressure.} To check whether there are more optimal charge
transfer and strain conditions than those in MS1 and MS2 we simulate
MS phases at other concentrations (Fig. 6, top panel). In the Li-B
system the MS3 and MS4 phases have energies well above the
$\alpha$-B$\leftrightarrow$$\alpha$-LiB and
$\alpha$-LiB$\leftrightarrow$fcc-Li tie-lines, so they will be
unstable against phase separation into the known compounds (these MS
phases remain metastable under hydrostatic pressure as well). This
test confirms our earlier finding that MS1-LiB and MS2-LiB are
particularly stable because of the near-optimal occupation of the
binding boron states\cite{MGB}.
\begin{figure}[t]
  \begin{center} \vspace{-5mm}
    \centerline{\epsfig{file=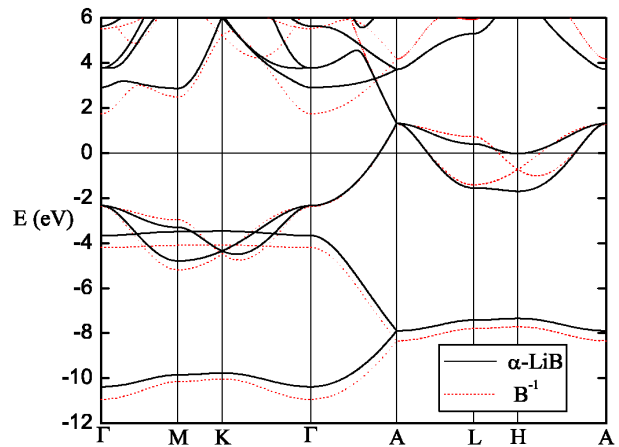,width=90mm,clip=}}
    \vspace{-8mm} \caption{ \small (color online). Band structures of
    $\alpha$-LiB (\cite{aLi}, solid black lines) and hypothetical
    B$^{-1}$ (described in the text, dotted red lines) phases. To
    facilitate the comparison of the band structures in the two phases
    we doubled the B$^{-1}$ unit cell along the $c$-axis and used the
    equilibrium lattice constants of $\alpha$-LiB. With the optimized
    $c$-axis the B$^{-1}$ electronic states in the -11 $\sim$ -2 eV
    range upshift by about 0.15 eV. Fermi level is at 0 eV.}
    \label{aLiB} \end{center}
\end{figure}

Our conjecture that the hydrostatic pressure would favor the MS
lithium monoboride phases over the off-stoichiometric LiB$_y$
compounds\cite{MGB} is also supported by the results shown in
Fig. 6. As far as the possibility of the MS-LiB formation is
concerned, it is rather unfortunate that the LiB$_y$ phases
additionally stabilize by becoming more lithium rich under
pressure. The information in Fig. 6 can be used to evaluate the
minimum pressure required to stabilize MS2-LiB with respect to
$\alpha$-B and LiB$_y$. We estimate by interpolation that at
$P_{min}\approx 5$ GPa the MS2-LiB phase lies on the line that goes
through $x=0$ and is tangent to the LiB$_y$ parabola. Therefore,
MS2-LiB could appear in the experiment at $P>P_{min}$ (not accounting
for the possible systematic errors and the thermodynamics
effects\cite{MGB}) if all other phases in the Li-B system were
metastable under such pressures. Using the parabola coefficients for
the three pressures given in Fig. 6 we also find by interpolation that
MS2-LiB has the same formation enthalpy as the most stable
LiB$_{y\approx 0.82}$ phase at about 12 GPa. Finally, we observe that
the line connecting MS2-LiB and fcc-Li crosses the parabolas in all
the cases for pressures below 30 GPa\cite{H}. This implies that if the
MS2-LiB phase was synthesized, LiB$_y$ might still be present in the
sample as a by-product. For analysis of the Li-B system at higher
pressures one needs to take into account that pure lithium undergoes
phase transformations from fcc to $hR1$ and eventually to $cI16$ near
40 GPa\cite{Li_pure}.

The chances for the formation of the MS-LiB phases depend on where
they are located on the phase diagram relative to LiB$_y$ and the most
lithium rich stable phase below 50\% concentration. The known phases
in this region have small atomic volume ($V_{Li_3B_{14}}=6.4$ and
$V_{LiB_3}=7.5$ \AA$^{3}$/atom under ambient
conditions\cite{Li3B14,LiB3,PF}) compared to MS2-LiB (11.2 and 6.7
\AA$^{3}$/atom at 0 and 30 GPa, respectively). The boron-rich phases
could potentially bar the formation of MS-LiB under pressure, however
they would need to have a very low formation enthalpy; for example, at
$P=12$ GPa it would need to be below $H_{MS2-LiB}= H_{LiB_{y\approx
0.82}}=$ -0.38 eV/atom. Simulation of a known phase with CaB$_6$
prototype, present in the K-B system\cite{KB6,Alkali}, could give
information on how boron-rich phases respond to hydrostatic
pressure. However, the ordered CaB$_6$-LiB$_6$ has a large atomic
volume and actually becomes less stable under pressure: $H_{f}=0.016$
eV/atom, $V=10.1$ \AA$^{3}$/atom at zero pressure and $H_{f}=0.303$
eV/atom, $V=8.6$ \AA$^{3}$/atom at 30 GPa pressure. Rosner and
Pickett pointed out that a compact pseudodiamond phase B32 (NaTl
prototype) might appear under pressure\cite{bLi}.  Using enthalpy
versus pressure curves for the $\alpha$-LiB and B32 phases we find the
crossover pressure to be 22 GPa. MS2-LiB stays below B32-LiB until
about 65 GPa. Overall, our simulations suggest that there might be a
window of pressures at which the MS-LiB phases can be synthesized.
\begin{figure}[t]
  \begin{center} \vspace{-5mm}
    \centerline{\epsfig{file=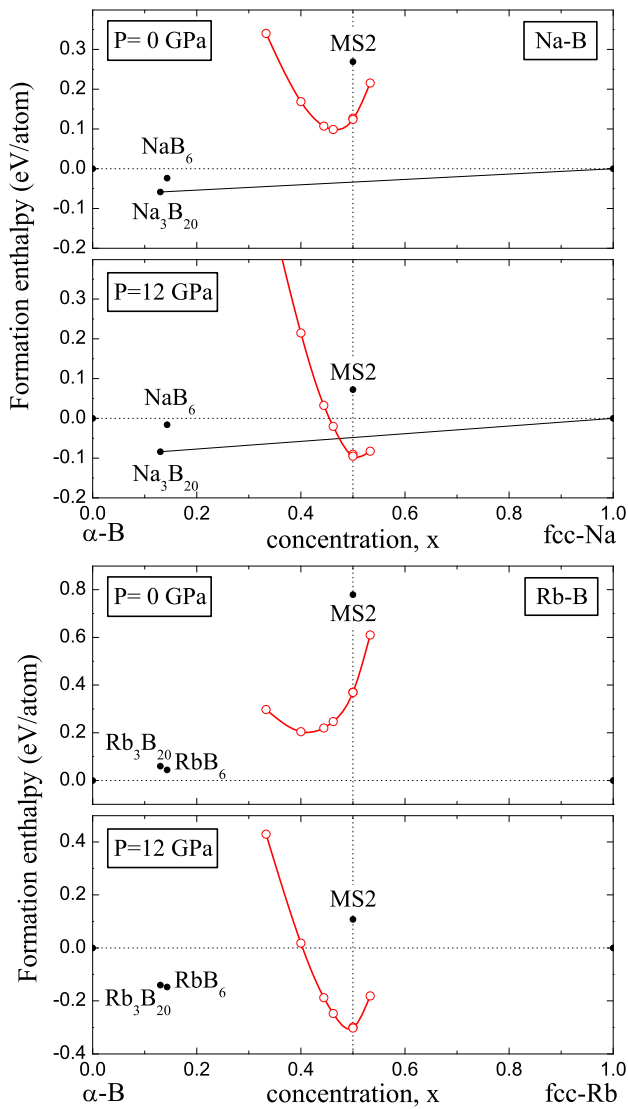,width=95mm,clip=}}
    \vspace{-8mm} \caption{ \small (color online). Calculated phase
    diagrams for Na-B and Rb-B systems at different pressures. Hollow
    red points are M$_{2n}$B$_m$ phases, from left to right:
    M$_2$B$_4$, M$_2$B$_3$, M$_4$B$_5$, M$_6$B$_7$, $\alpha$-MB,
    $\beta$-MB, and M$_8$B$_7$. The boron rich phases are
    Na$_3$B$_{20}$ and CaB$_6$ prototypes with Pearson symbols $oS46$
    and $cP7$, respectively. The vertical dotted lines mark the 50\%
    composition.}
    \label{phd13} \end{center}
\end{figure}
\begin{figure}[t]
  \begin{center} \vspace{-5mm}
    \centerline{\epsfig{file=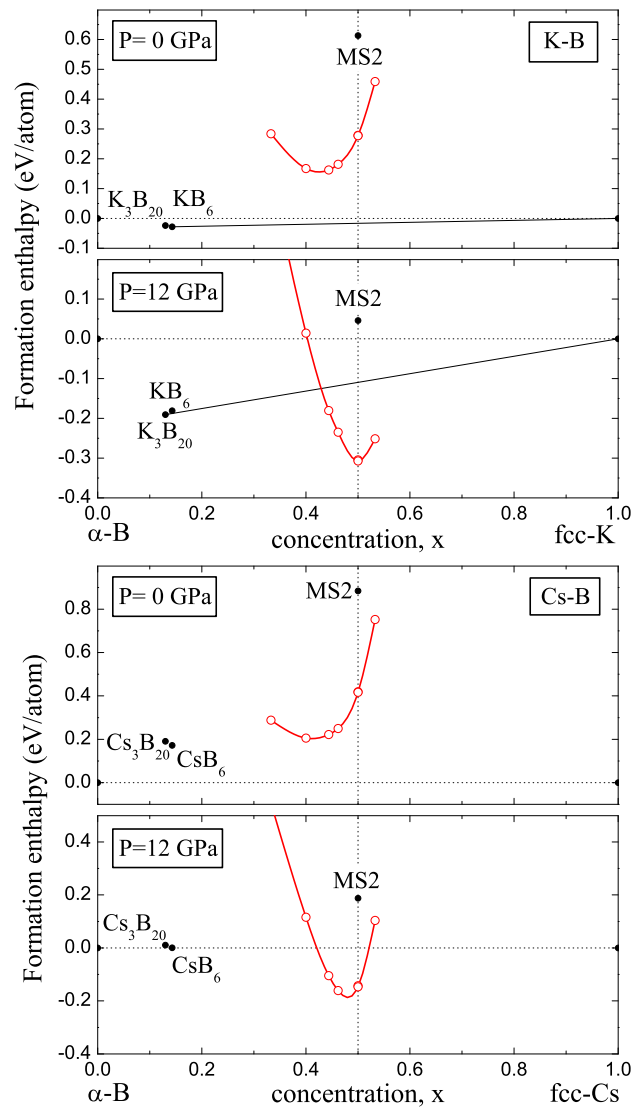,width=95mm,clip=}}
    \vspace{-8mm} \caption{ \small (color online). The same as in
    Fig. 11 for K-B and Cs-B systems.}
    \label{phd24} \end{center}
\end{figure}

\section{6. Other monovalent and some higher-valent metal borides}
\label{section.Noble}

{\it Alkali and transition metal borides.} According to the
experimental databases and the latest review of alkali-metal
borides\cite{PF,Alkali} there are no stable sodium or potassium
borides above 15\% concentration of metal and no stable rubidium or
cesium borides in the whole concentration range. Na$_3$B$_{20}$
(Pearson symbol $oS46$) and KB$_6$ (Pearson symbol $cP7$) compounds,
made out of boron polyhedra intercalated with alkali atoms, are
considered the most metal-rich known borides in the Na-B and K-B
systems, respectively\cite{Na3B20,KB6,PF,Alkali}. Our simulations
confirm that these compounds have negative formation enthalpies of -58
meV/atom for Na$_3$B$_{20}$ and -29 meV/atom for KB$_6$ with respect
to $\alpha$-B and fcc-M\cite{bcc}. The formation Gibbs free energies
for these compounds might be less negative if they were evaluated with
respect to $\beta$-B at finite temperature\cite{bB}. This could be the
reason why there is no conclusive evidence of potassium hexaboride
synthesis\cite{Alkali}. Complete theoretical investigation of the
boron-rich compounds such as Na$_3$B$_{29}$ (Ref. \cite{Na3B29}) is
beyond the scope of this study but it is interesting to see where the
nearly stoichiometric MB$_y$ phases place with respect to the known
phases in these systems.

As in the LiB$_y$ compounds, the metal and boron sublattices in the
alkali boride phases MB$_y$ (M = Na, K, Rb, Cs) are found to be very
weakly correlated. In fact, the larger alkali atoms push the boron
chains farther apart (see Table I) weakening the boron interchain
bonds (note the lower energy difference between $\alpha$-MB and
$\beta$-MB listed in Table I). This again leads to the situation when
MB$_y$ (M = Na, K, Rb, Cs) compounds can easily adapt to an optimal
composition by having incommensurate metal and boron
sublattices. Figures 11 and 12 demonstrate that in all Na-B, K-B,
Rb-B, Cs-B systems the stoichiometric phases with linear chains of
boron prefer to lose some metal, the opposite tendency compared to the
Li-B system. While the most stable LiB$_y$ composition appears to be
determined primarily by the optimal level of boron doping, in the
larger alkali-metal borides, MB$_y$, the lattice mismatch between the
metal and boron sublattices must be playing a more significant
role. Note that the formation enthalpy points for these M$_{2n}$B$_m$
phases are not symmetric, curving up more rapidly in the metal rich
region.

The stabilization from losing a few percent of alkali metal is
noticeable but not enough for NaB$_y$ and KB$_y$ to have a negative
formation enthalpy at zero pressure. Because the interchain spacing is
determined mostly by the alkali cations the $C_{11}+C_{12}$ force
constant in $\beta$-MB decreases as one moves down the periodic table
(see Table I). $C_{33}$ also becomes smaller as the boron-boron bond
length gets longer. The softness of the MB$_y$ phases (M = Na, K, Rb,
Cs) invites the use of hydrostatic pressure for their synthesis.
Moreover, in the Na-B and K-B systems the MB$_y$ phases stabilize more
rapidly than the MS2-MB phases, which makes it unlikely for the latter
to form under the pressures considered. Synthesis of MB$_y$ (M = Na,
K, Rb, Cs) would provide valuable information on the ways the linear
chains of boron could be stabilized. Because the alkali-metal borides
are not fully explored, it would not be surprising if a not considered
here or a completely unknown phase appeared in such an experiment.
\begin{table}[b]
  \begin{center}
  \caption{ \small Calculated properties of $\beta$-MB metal borides
  (NiAs prototype).}
  \vspace{-1mm}
  \begin{tabular}{c|ccccc}
 \hline\hline
  compound           &     $a_0$  &   $c_0$    &$C_{11}+C_{12}$&  $C_{33}$   & $E_{\beta-MB}-E_{\alpha-MB}$\\
                     &     (\AA)  &   (\AA)    &   (GPa)       &   (GPa)     &      (meV/atom)             \\
\hline
  LiB                &    4.013   &   3.120    &   139         &   542       &        -10                  \\
  NaB                &    4.697   &   3.196    &   111         &   379       &        -3.1                 \\
   KB                &    5.390   &   3.240    &    79         &   268       &        -1.6                 \\
  RbB                &    5.662   &   3.267    &    77         &   251       &        -1.7                 \\
  CsB                &    5.976   &   3.325    &    69         &   252       &        -2.0                 \\
\hline
  RhB                &    3.382   &   4.185    &   498         &   309       &       -267                  \\
  PtB                &    3.765   &   3.655    &   546         &   293       &       -108                  \\
\hline\hline
\end{tabular}
\label{NiAs}
\end{center}
\end{table}

While the $\beta$-MB phases are only metastable for the borides in the
alkali-metal series, there are two reported stable transition-metal
monoborides in this configuration: RhB and PtB
(Ref. \cite{PF,RhB_PtB}). Our fully relaxed unit cell parameters (see
Table I) agree well with experiment for $\beta$-RhB ($a=3.309$ \AA,
$c=4.224$ \AA), but they disagree by over 10\% with the measured
values ($a=3.358$ \AA, $c=4.058$ \AA) for
$\beta$-PtB\cite{PF,RhB_PtB}. Identification of the source of this
discrepancy requires additional study of this system. Nevertheless,
the data on $\beta$-RhB and $\beta$-PtB in Table I give an idea about
what difference the $d$-electrons cause in the boron-boron binding
compared to the case of the alkali-metal monoborides. For example,
$\beta$-RhB and $\beta$-PtB no longer have the optimally doped
double-bonded boron chains: the boron-boron bond is so overstretched
that it exceeds the $sp^2$ bond length in the AlB$_2$-type compounds,
resulting in the increase of the $c$-axis compressibility compared to
the alkali-metal monoborides. The significant reduction of the
interchain distances leads to an over 300\% increase in the
$C_{11}+C_{12}$ force constants. One more important consequence of the
more compact arrangement of atoms in the lateral direction and the
hybridization of the $d$-orbitals of metal with the valence states of
boron is the much larger energy difference between the $\beta$-MB and
$\alpha$-MB structures. This makes the formation of the
off-stoichiometry phases with linear chains of boron in the
transition-metal monoborides energetically unfavorable.
\begin{figure}[t]
  \begin{center} \vspace{-5mm}
    \centerline{\epsfig{file=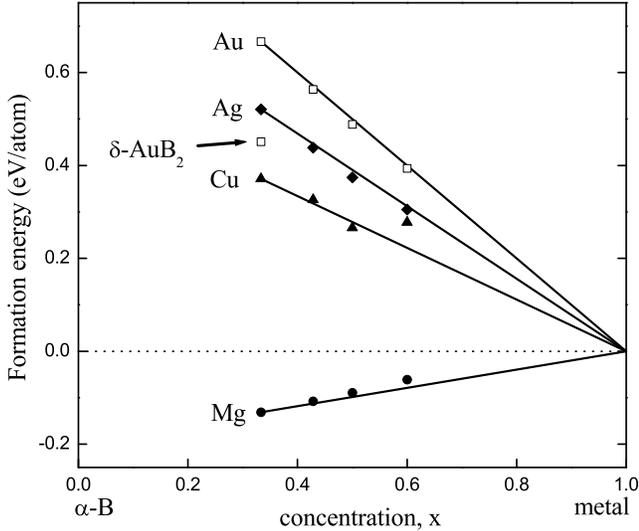,width=95mm,clip=}}
    \vspace{-8mm} \caption{ \small (color online). Calculated phase
    diagrams for magnesium and noble-metal borides. The solid points
    are layered phases from left to right: $|\alpha|$, MS4, MS2, MS3.
    We also show $\delta$-AuB$_2$ phase\cite{deltaAu}, which is
    significantly lower than $|\alpha|$-AuB$_2$.}
    \label{noble} \end{center}
\end{figure}

{\it Noble- and divalent-metal borides.} AgB$_2$ and AuB$_2$ have been
shown to have big positive formation energies\cite{Oguchi}, so they
should not form at ambient pressure. Recent experiments suggest that
some superconducting Ag-B phase was formed by pulsed laser
deposition\cite{Ag_laser}. Because of the synthesis conditions, the
thin-film samples were inhomogeneous and did not produce new x-ray
peaks. It was assumed that the observed phase was
$|\alpha|$-prototype, although the $T_c$ turned out to be much lower
than the anticipated value\cite{hype}. Formation of Ag vacancies in
$|\alpha|$-Ag$_x$B$_2$ was suggested by Shein {\it et al.} as a
possible explanation of the observed data\cite{AgxB2}. Current
simulations offer other possibilities for formation of phases under
non-equilibrium conditions: the proposed phases still have positive
formation energies but they are less unstable and below the respective
$|\alpha|\leftrightarrow$metal tie-lines (Fig. 13). The intermediate
phases in the concentration range from 33\% to 60\% generally stay
below the tie-lines. If formed, MS2-AgB would be even more dynamically
stable than MS2-LiB: we find that the former represents a stable
equilibrium with the frequencies of the softest interlayer modes
$\omega_{x,y}$ three times higher than those in the latter\cite{MGB}.
This is an expected result, considering that the silver bilayer in
MS2-AgB remains bound by the $d$-electrons even if it donates most of
the charge from the $s$-orbital to boron as lithium does. The
electronic properties and the stability of the MS noble-metal borides
are further discussed in Section 7.

We find that the $\delta$-MB$_2$ phase\cite{deltaAu} is surprisingly
much more stable than the $|\alpha|$-MB$_2$ phase for several metals:
Au, Ca, K, Pd, and Pt (by 215, 28, 214, 28, and 272 meV/atom,
respectively). Apparently, metal atoms prefer to hybridize their
valence states with $p$-orbitals of boron more strongly by shifting to
a boron-boron bond, rather than simply donate their valence
electrons. Note that the five metals are either large in size
($d_{Ca}$=3.88 \AA, $d_K$=4.71 \AA, see Fig. 4) or have a big work
function (bulk values: $\phi_{Au}$=5.1 eV, $\phi_{Pd}$=5.12 eV,
$\phi_{Pt}$=5.65 eV\cite{WF,monolayer}). The discovery of the
lower-symmetry $\delta$-AuB$_2$ phase\cite{deltaAu} rules out the
possibility of $|\alpha|$-AuB$_2$ synthesis. The formation energy of
$\delta$-AuB$_2$ remains positive (0.45 eV/atom, see Fig. 13), which
makes this phase unlikely to form as well. In the Ca-B system
$\delta$-CaB$_2$ is still unstable against phase separation into
CaB$_6$ and fcc-Ca \cite{Oguchi,PF} by 144 meV/atom.

The MS phases in the Mg-B system stay at least a few meV/atom above
the $|\alpha|$-MgB$_2$$\leftrightarrow$hcp-Mg tie-line. Hydrostatic
pressure is insignificant to their relative stability because they
compete against similar phases. Therefore, the MS magnesium boride
phases are not likely to form and could possibly exist only in the
form of a defect in $|\alpha|$-MgB$_2$. We calculate the following
series: $|\alpha\beta|$, $|\alpha|\alpha\beta|$,
$|\alpha|\alpha|\alpha\beta|$, and so on up to 8 $|\alpha|$ unit
cells, and find by extrapolation that the energy required to insert a
single magnesium layer into the $|\alpha|$-MgB$_2$ matrix is quite
high: over 25 meV per atom in the additional layer of magnesium.

\section{7. Summary}
\label{section.summary}

\begin{table}[b]
  \begin{center}
  \caption{ \small Metal boride phases: formation energy ($E_{f}$,
    eV/atom), relative stability ($\Delta E_X$, eV/atom, Eq. 1),
    in-plane boron-boron bond ($a_{B-B}$, \AA), interplanar
    boron-boron distance ($c_{B-B}$, \AA), internal coordinate of the
    metal atom ($z_M$)\cite{MS2,MS3}, position of $p\sigma$-band in
    boron at $\Gamma$ and $A$ k-points ($E_{\Gamma}$, $E_A$, eV), and
    PDOS at $E_F$ for B-$p_{xy}$ ($N_{p_{xy}}^B(0)$,
    states/(eV$\cdot$spin$\cdot$ boron atom)).}
  \vspace{0mm}
  \begin{tabular}{cc|ccccc}
 \hline\hline
  phase              &                     &     Li     &    Ag    &    Au    &    Cu    &       Mg     \\
\hline
                     &      $E_{f}$        &   0.003    &  0.520   &  0.667   &  0.371   &    -0.131    \\
                     &     $a_{B-B}$       &   1.717    &  1.744   &  1.737   &  1.721   &     1.776    \\
   $|\alpha|$        &     $c_{B-B}$       &   3.469    &  4.081   &  4.260   &  3.382   &     3.521    \\
    (33\%)           &  $E_{\Gamma}$       &   1.48     &  1.29    &  1.26    &  1.14    &     0.39     \\
                     &  $E_A$              &   1.70     &  1.01    &  0.95    &  0.65    &     0.77     \\
                     &  $N_{p_{xy}}^B(0)$  &   0.076    &  0.090   &  0.093   &  0.106   &     0.049    \\
\hline
                     &  $E_{f}$            &  -0.162    &  0.374   &  0.488   &  0.266    &   -0.089    \\
                     &  $\Delta  E_{MS2}$  &  -0.164    & -0.016   & -0.012   & -0.013    &    0.009    \\
    MS2              &   $a_{B-B}$         &   1.765    &  1.739   &  1.734   &  1.734    &    1.805    \\
 (50\%)              &   $c_{B-B}$         &   5.522    &  6.369   &  6.589   &  4.915    &    5.989    \\
   \cite{MS2}        &   $z_M$             &   0.496    &  0.368   &  0.356   &  0.388    &    0.440    \\
                     &  $E_{\Gamma}$       &   0.99     &  1.19    &  1.17    &  0.69     &    0.15     \\
                     &  $E_A$              &   0.99     &  1.19    &  1.17    &  0.69     &    0.15     \\
                     &  $N_{p_{xy}}^B(0)$  &   0.059    &  0.086   &  0.091   &  0.091    &    0.043    \\
\hline
                     &  $E_{f}$            &  -0.117    &  0.305   &  0.394   &  0.277    &   -0.061    \\
                     &  $\Delta  E_{MS3}$  &  -0.118    & -0.007   & -0.006   &  0.054    &    0.018    \\
    MS3              &   $a_{B-B}$         &   1.745    &  1.731   &  1.722   &  1.695    &    1.816    \\
 (60\%)              &   $c_{B-B}$         &   8.318    &  8.785   &  9.097   &  6.903    &    8.511    \\
   \cite{MS3}        &   $z_M$             &   0.178    &  0.231   &  0.237   &  0.228    &    0.194    \\
                     &  $E_{\Gamma}$       &   1.23     &  1.22    &  1.25    &  0.88     &    0.03     \\
                     &  $E_A$              &   1.23     &  1.22    &  1.25    &  0.89     &    0.03     \\
                     &  $N_{p_{xy}}^B(0)$  &   0.066    &  0.090   &  0.098   &  0.074    &    0.039    \\
\hline\hline
\end{tabular}
\label{table2}
\end{center}
\end{table}

{\it Electronic properties.} It is illustrative to compare the
important features of the electronic structure in the low-valent metal
borides to those in the lithium borides, which were discussed in our
previous study\cite{MGB}. We focus mainly on the $p$-states of boron,
important for the stability and the superconductivity in these
compounds. For convenience, we calculate the band structure and PDOS
for $|\alpha|$, MS2, and MS3 phases because all three have a hexagonal
unit cell. The key characteristics of the boron states, along with
parameters of the unit cells for these phases are given in Table II.

For all MS2 metal borides the $p\sigma$ band along $\Gamma$-A is
practically flat because of the large separation between boron layers,
as shown in Fig. 4 of Ref.\cite{MGB} for MS2-LiB and in Fig. 14 for
MS2-AgB. For Ag and Au this band does not move much from the
respective average positions in $|\alpha|$ and the PDOS of $p\sigma$
states in boron at the Fermi level, $N_{p_{xy}}^B(0)$, stays nearly
the same (Table II and Fig. 15). In fact, these boron properties
remain the same in the more metal-rich structure MS3. The results
suggest that the level of doping of the boron layers in the MS noble
metal borides is nearly independent of the number of metal
layers. Considering this and the fact that the MS phases in the
33\%-100\% range closely follow the $|\alpha|\leftrightarrow$fcc lines
on the Ag-B and Au-B phase diagrams these MS phases can be viewed as a
mixture of weakly interacting building blocks: the $|\alpha|$ unit
cell with an established charge redistribution within it and the
closed-packed layers of pure metal. Therefore, one could expect the
superconducting properties of the boron layer in the hypothetical
phases $|\alpha|$-AgB$_2$ and the MS silver borides in the 33\%-100\%
concentration range to be similar. However, the nonequilibrium
conditions necessary to synthesize such compounds\cite{Ag_laser} may
introduce disorder destroying the superconducting states\cite{LixBC}.

Compared to the MS gold and silver borides, the MS copper borides
deviate more from the $|\alpha|$-CuB$_2$$\leftrightarrow$fcc-Cu line
(Fig. 13) and the level of boron doping is more nonmonotonic as a
function of the metal concentration (Table II), even though the work
functions of Cu and Ag are close (bulk values: $\phi_{Cu}$ = 4.65 eV,
$\phi_{Ag}$ = 4.26 eV, Ref. \cite{WF}). The different behavior must be
the result of the more pronounced strain between the copper and boron
layers caused by the smaller size of copper ($d_{Cu}$ = 2.563 \AA,
$d_{Ag}$ = 2.937 \AA, $d_{Au}$ = 2.949 \AA, see Fig. 4). In the more
electron-rich Mg-B system the level of boron doping increases as
additional magnesium layers are added. In MS3-Mg$_3$B$_2$ the
$p\sigma$ states are almost completely occupied.

Lithium turns out to be a special case among the monovalent metals: it
has the right size and can easily donate electrons to stabilize the
boron layers in the MS phases. Fig. 6 and Table II show that the
stabilization effect is largest for the 50\% concentration MS1 and MS2
lithium borides. No less importantly, by giving up most of their
charge the lithium layers interact only weakly with one another, which
makes the compound very soft along the $c$-axis and gives the
opportunity to stabilize the compound even more with hydrostatic
pressure\cite{MGB}. The addition of extra layers does not result in
further stabilization, as boron in MS3-Li$_3$B$_2$ is doped less than
in MS2-LiB, judging by the location of the $p\sigma$ band along
$\Gamma$-A (Table II and Fig. 15). The present PAW calculations
confirm our previous result that MS2-LiB has a higher PDOS of boron
$p\sigma$ states at the Fermi level than that in MgB$_2$
(Ref. \cite{MGB}). Note that the PDOS in our simulations is found by
decomposition of the wavefunction within a sphere of fixed radius and
can slightly vary with this parameter, as well as with the
approximation used. In the APW+lo calculation we obtained a 12\%
increase of $N_{p_{xy}}^B(0)$ in MS2-LiB compared to that in MgB$_2$
using $R^B_{MT}$ = 1.6 a.u.\cite{MGB}, while in the present PAW
calculations we observe a 20\% increase using the default PAW radius
of 1.7 a.u.
\begin{figure}[t]
  \begin{center} \vspace{-5mm}
    \centerline{\epsfig{file=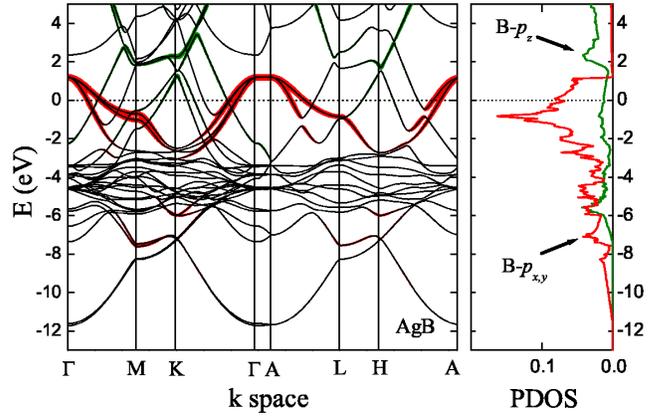,width=95mm,clip=}}
    \vspace{-8mm} \caption{ \small (color online). Band structure and
    partial density of states (PDOS) in a hypothetical MS2-AgB phase,
    calculated in APW+lo\cite{WIEN2K,lapw}. PDOS units are
    states/(eV$\cdot$spin) per boron atom. The thickness of band
    structure lines is proportional to boron $p_{x,y}$ (red) and $p_z$
    (green) characters. Fermi level is at 0 eV.}
    \label{AgB} \end{center}
\end{figure}
\begin{figure}[t]
  \begin{center} \vspace{-5mm}
    \centerline{\epsfig{file=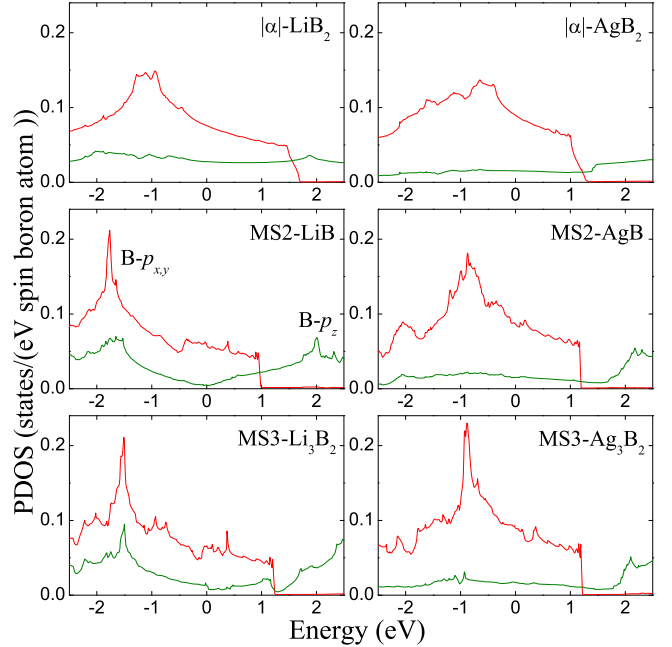,width=95mm,clip=}}
    \vspace{-8mm} \caption{ \small (color online). PDOS of boron
    $p_{x,y}$ (red) and $p_z$ (green) states in lithium and silver MS
    borides calculated with {\small VASP}. Fermi level is at 0 eV.}
    \label{psDOS} \end{center}
\end{figure}

The ample amount of the boron $p\sigma$ states at the Fermi level in
MS2-LiB holds great promise for this compound to be a good
superconductor. However, a more thorough calculation of the
electron-phonon coupling, such as in Ref. \cite{Calandra}, is required
to say with certainty whether the new lithium monoboride can compete
with the record-holder MgB$_2$. Such calculation is
underway\cite{MGC}.

{\it X-ray reflections.} As we pointed out previously\cite{MGB}, the
resulting structure in metal borides at 1:1 composition could be a
random mixture of MS1 and MS2, because they differ only by a
long-period shift in stacking order and, therefore, are nearly
degenerate. The two phases also have very close nearest neighbor
distances in most systems and random structures would still have a
constant separation between boron layers. For magnesium and silver MS
monoborides these periods correspond to $2\theta=16.4^\circ$ and
$2\theta=16.8^\circ$, respectively (for $\lambda=1.5418$ \AA).

The largest difference (3\%) in the c-axis for MS1 and MS2 is actually
found for Li-B, resulting in $2\theta=16.6^\circ$ (d$_{MS1-LiB}=5.35$
\AA) and $2\theta=16.1^\circ$ (d$_{MS2-LiB}=5.52$ \AA); the more
lithium-rich phase MS3-Li$_3$B$_2$ would produce a reflection at
$2\theta=10.6^\circ$ (d$_{MS3-Li_3B_2}=8.32$ \AA). The most pronounced
peaks in the published x-ray data for lithium monoboride are at
$2\theta$ = $25.5^\circ$, $41.3^\circ$, and $45.0^\circ$ which fit
well to the calculated $\alpha$-LiB x-ray pattern\cite{aLi}.
Interestingly, two reflections at low angles $2\theta$ = $12.2^\circ$
and $20.9^\circ$ were observed at 40-50\% of lithium
concentration\cite{Wan78}. A low-angle reflection was also detected at
12.8$^\circ$ in Al$_{1-x}$Li$_x$B$_2$ under heavy Li
doping\cite{AlLiB}. However, none of the observed peaks in the samples
prepared at ambient pressure match the calculated x-ray reflections in
the MS lithium borides.

\section{8. Conclusions}
\label{section.conclusions}
The main results of the present study can be summarized as follows:

{\bf i.} We have identified a previously unknown class of metal-rich
   layered phases that are comparable in energy to existing metal
   borides. This interesting accidental result should be credited to
   the exhaustive consideration of all candidates in the DMQC method
   and the careful structural relaxation in the calculation of their
   ground state energies.

{\bf ii.} Our {\it ab initio} results suggest that the MS phases are
   most suitable for electron-deficient metal boride systems. In the
   Ag-B and Au-B systems the MS phases are less unstable than the
   corresponding diborides with the AlB$_2$ prototype but they still
   have positive formation energies. In the Mg-B system the MS phases
   are metastable and could possibly exist only as a defect in
   MgB$_2$.

{\bf iii.} The MS-LiB phases present a special case among the MS metal
   borides: lithium has the right size and valence to stabilize the
   hexagonal layers of boron at 1:1 composition. The MS lithium
   monoboride phases are shown to have lower formation enthalpy with
   respect to the experimentally observed nearly stoichiometric
   LiB$_y$ phases under hydrostatic pressure. This encouraging result
   suggests that the new superconducting MS-LiB phases might form
   under proper conditions. The lowest required pressure depends on
   the position of boron-rich phases in the Li-B phase diagram and
   could be as low as several GPa.

{\bf iv.} For a more complete description of the Li-B system we
   introduce a simple model of the off-stoichiometric LiB$_y$ phases
   which explains the available experimental data. We demonstrate that
   because of the weak correlation between the boron and lithium
   sublattices the compound can easily adapt to an optimal
   composition, which corresponds to an optimal level of boron doping
   with the Fermi level lying near the bottom of the
   pseudogap. Interestingly, these phases turn out to be stable over a
   {\it range} of concentrations around $y=0.9$, in excellent
   agreement with experiment. We list the relaxed unit cell parameters
   which should be helpful in the determination of the LiB$_y$
   composition.

{\bf v.} We consider the MB$_y$ phases for other alkali-metal borides
   and find that these compounds also benefit from going
   off-stoichiometry, only in this case they prefer to lose some
   metal. The ensuing gain in enthalpy does not make them stable under
   ambient conditions, however these phases might form under
   hydrostatic pressure. Synthesis of the MB$_y$ phases (M = Na, K,
   Rb, Cs) would provide valuable information on how linear chains of
   boron could be stabilized.

We thank M. Calandra, F.H. Cocks, V. Crespi, P. Lammert, E. Margine,
and J. Sofo for valuable discussions. We acknowledge the San Diego
Supercomputer Center for computational resources.




\begin{thebibliography}{99}
\bibitem{xy}
To facilitate comparison with experiment we use the LiB$_y$ notation
for boron-deficient lithium monoboride phases. Concentration of boron
is denoted as $y$ instead of $x$ to avoid confusion with the
alternative notation Li$_x$B$_{1-x}$ ($y=(1-x)/x$).
\bibitem{DMQC}
Description of the methods and calculation details can be found in
\cite{SC1}.
\bibitem{bcc}
Li, Na, K, Rb, and Cs have nearly degenerate bcc and fcc energies at
zero pressure.
\bibitem{stress}
The difference of the equilibrium boron-boron bond lengths in the
LiB$_y$ and B$^{1/y}$ phases caused by the constant stress from the
lithium sublattice ($F_{Li^+}=0.76$ eV/\AA) is found as
$a_{LiB_y}-a_{B^{-1/y}}$=$-F_{Li^+}/( \partial^2
E_{B^{-1/y}}(a)/\partial a^2)$, where $\partial^2
E_{B^{-1/y}}(a)/\partial a^2 $ is 26.5 eV/\AA$^2$ at y=1 and 22.8
eV/\AA$^2$ at y=0.8.
\bibitem{H} 
In fact, if we define $H^*$ as the intersection of the tangent and the
$x=0.5$ line in Fig. 6 ($H^*\equiv -0.5\times tangent$), then $H^*$ is
lower than $H_{MS2-LiB}$ by approximately 40 meV/atom for all the
pressures considered.
\bibitem{monolayer} 
Work functions of a [111] monolayer of different metals are likely to
follow the same order as those of bulk structure.
\end{thebibliography}
\end{document}